\documentclass[runningheads]{llncs}
\usepackage[T1]{fontenc}

\usepackage{graphicx}
\usepackage{color}
\usepackage{amsmath,amssymb,amsfonts}
\usepackage{pifont}
\usepackage{multirow}
\usepackage{makecell}
\usepackage{bm}
\usepackage{bbm}
\usepackage{booktabs}
\usepackage[misc]{ifsym}

\definecolor{myRed}{RGB}{234,51,35}
\definecolor{myGreen}{RGB}{79,173,91}
\definecolor{myOrange}{RGB}{178,86,40}
\definecolor{myBlue}{RGB}{51,94,150}
\definecolor{myYellow}{RGB}{245,194,66}
\definecolor{myPurple}{RGB}{120,32,110}

\definecolor{WorkflowBlue}{RGB}{70,177,225}
\definecolor{WorkflowGreen}{RGB}{114,209,105}
\definecolor{WorkflowGray}{RGB}{116,116,116}

\usepackage{hyperref}
\hypersetup{colorlinks = True, 
	    linkcolor = red, 
        citecolor = blue} 

\begin{document}
\title{FastSAM-3DSlicer: A 3D-Slicer Extension for 3D Volumetric Segment Anything Model with Uncertainty Quantification}
\titlerunning{FastSAM-3DSlicer}
\author{
Yiqing Shen\inst{1} \and Xinyuan Shao\inst{1} \and Blanca Inigo Romillo\inst{1} \and David Dreizin\inst{2} \and Mathias Unberath\inst{1}\textsuperscript{(\Letter)}
}
\authorrunning{Y. Shen et al.}
\institute{
Johns Hopkins University, Baltimore, MD, 21218, USA \and
University of Maryland School of Medicine and R Adams Cowley Shock Trauma Center, Baltimore, MD, 21201, USA\\
\email{\{yshen92,unberath\}@jhu.edu}
}

\maketitle              
\begin{abstract}
Accurate segmentation of anatomical structures and pathological regions in medical images is crucial for diagnosis, treatment planning, and disease monitoring. 
While the Segment Anything Model (SAM) and its variants have demonstrated impressive interactive segmentation capabilities on image types not seen during training without the need for domain adaptation or retraining, their practical application in volumetric 3D medical imaging workflows has been hindered by the lack of a user-friendly interface.
To address this challenge, we introduce FastSAM-3DSlicer, a 3D Slicer extension that integrates both 2D and 3D SAM models, including SAM-Med2D, MedSAM, SAM-Med3D, and FastSAM-3D. 
Building on the well-established open-source 3D Slicer platform, our extension enables efficient, real-time segmentation of 3D volumetric medical images, with seamless interaction and visualization.
By automating the handling of raw image data, user prompts, and segmented masks, FastSAM-3DSlicer provides a streamlined, user-friendly interface that can be easily incorporated into medical image analysis workflows.
Performance evaluations reveal that the FastSAM-3DSlicer extension running FastSAM-3D achieves low inference times of only 1.09 seconds per volume on CPU and 0.73 seconds per volume on GPU, making it well-suited for real-time interactive segmentation. 
Moreover, we introduce an uncertainty quantification scheme that leverages the rapid inference capabilities of FastSAM-3D for practical implementation, further enhancing its reliability and applicability in medical settings.
FastSAM-3DSlicer offers an interactive platform and user interface for 2D and 3D interactive volumetric medical image segmentation, offering a powerful combination of efficiency, precision, and ease of use with SAMs.
The source code and a video demonstration are publicly available at  \url{https://github.com/arcadelab/FastSAM3D_slicer}.
\end{abstract}

\section{Introduction}
Precise segmentation of anatomical structures and pathological regions from medical images is essential for accurate diagnosis, treatment planning, and monitoring of disease progression \cite{nnunet}. 
However, manual segmentation is time-consuming, labor-intensive, and prone to inter-observer variability.
Deep-learning-driven automatic segmentation models have shown promise in reducing manual effort, they often struggle to generalize across diverse datasets, anatomical variations, and unseen pathologies during inference \cite{liu2021review}.
The Segment Anything Model (SAM) \cite{sam} and its variants for volumetric medical images, such as SAM-Med3D \cite{sammed3d}, have emerged as flexible zero-shot solutions that enable interactive segmentation of novel objects without requiring retraining.
These models are designed to generalize across various tasks and datasets through large-scale pre-training and support for manual prompting.
The interactive nature of SAM-based segmentation makes fast inference times particularly important, as it allows users to make immediate adjustments and corrections, improving the accuracy and efficiency of the segmentation process.
FastSAM-3D \cite{shen2024fastsam3d}, a computationally efficient 3D SAM architecture, has been specifically optimized for real-time interactive segmentation of 3D volumes such as computed tomography (CT). 
FastSAM-3D utilizes a compact Vision Transformer (ViT) encoder \cite{vit}, distilled from the larger SAM-Med3D \cite{sammed3d}, and incorporates an efficient 3D Sparse Flash Attention \cite{shen2024fastsam3d} mechanism to reduce computational costs while maintaining high segmentation quality.

Despite the potential of efficient SAM variants to enable highly responsive interactive segmentation, there is currently a lack of user-friendly interfaces to facilitate their practical application in medical image analysis workflow. 
Interactive prompting in 3D medical volumes is more challenging compared to 2D interfaces for natural images, due to the increased complexity of visualizing 3D data on a 2D screen and the need for seamless integration with existing medical image analysis workflows. 
3D Slicer\footnote{\url{https://www.slicer.org}} \cite{3dslicer} is an open-source software platform widely used for the analysis and visualization of volumetric medical images. 
It supports a variety of plug-ins and offers a robust framework for the integration of new tools, making it an ideal choice for creating an interface for interactive volumetric image segmentation, e.\,g., with FastSAM-3D.
In this manuscript, we describe FastSAM-3DSlicer, a plugin for integrating 2D and 3D SAM models, including the efficient FastSAM-3D, into the well-established image analysis platform 3D Slicer. 
This extension enables users to load 3D volumes, interactively annotate structures of interest using point prompts, and visualize the resulting segmentations in real time within a familiar software environment.

In summary, our contributions are three-fold.
Firstly, we propose a novel 3D Slicer-based extension for both 2D and 3D SAM models, including the efficient FastSAM-3D, for volumetric image segmentation.
It enables interactive prompting in a 3D manner with SAM.
%
%
Secondly, we show the quantitative comparison of the inference time of different SAM models within our interface on both CPU and GPU environments. 
Finally, we propose an uncertainty quantification scheme based on the fast inference speed of FastSAM-3D in our extension.
It can guide the user for better prompting.

\begin{figure}[ht!]
    \centering
    \includegraphics[width=0.92\linewidth]{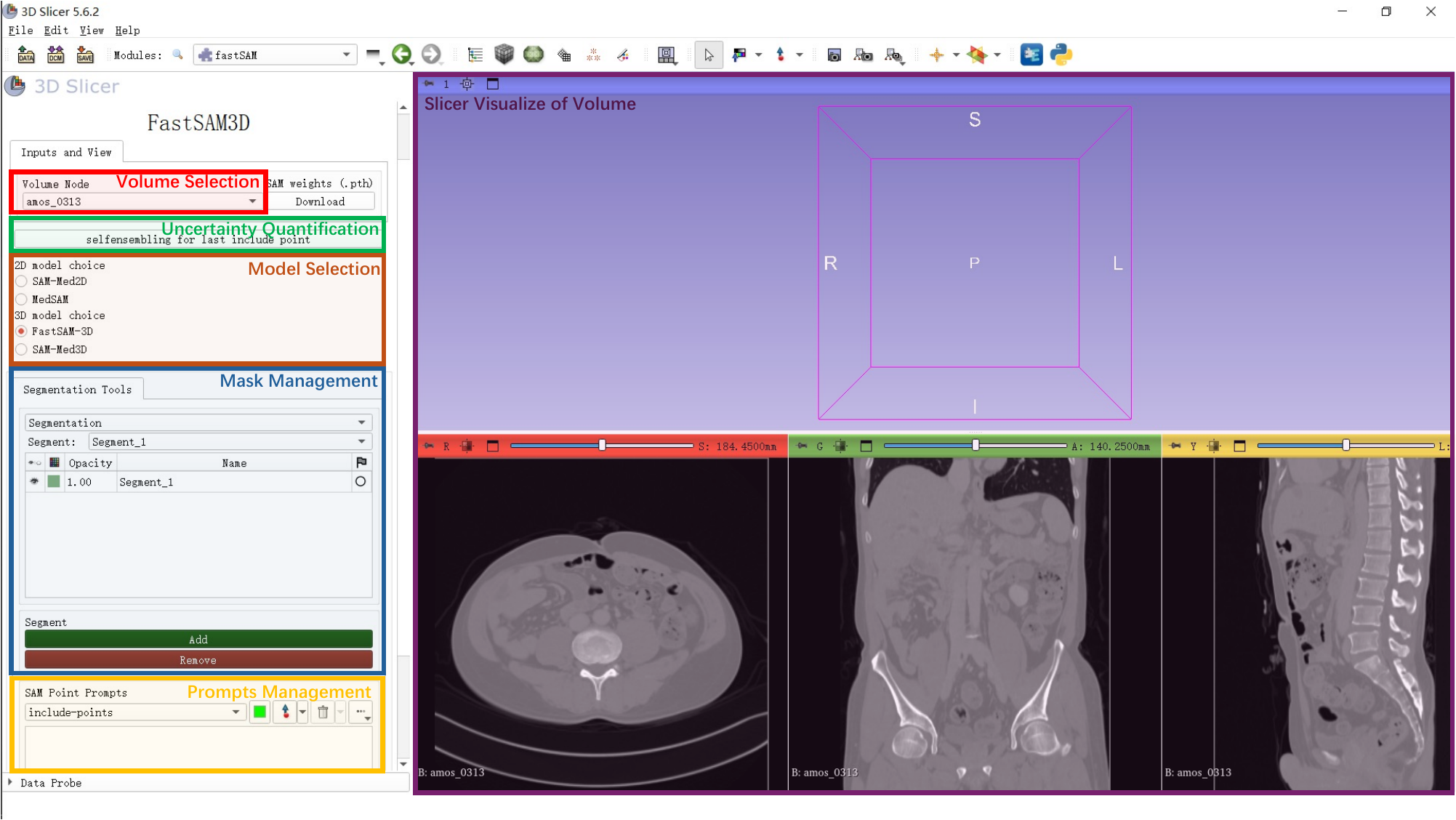}
    \caption{
An illustration of the interface of our 3D Slicer extension \textit{i}.\textit{e}., FastSAM-3DSlicer. 
It allows users to load 3D volumes (\textcolor{myRed}{red}), select the mode for uncertainty quantification (\textcolor{myGreen}{green}), choose SAM models (\textcolor{myOrange}{brown}), manage masks (\textcolor{myBlue}{blue}) and prompts (\textcolor{myYellow}{yellow}), and visualize segmentation and input volume in real-time (\textcolor{myPurple}{purple}).
%
%
This integration enhances image analysis workflows by providing an intuitive and efficient interface for the interactive annotation and segmentation of volumetric medical images with SAM.
    }\label{fig:interface}
\end{figure}

\begin{figure}[ht!]
    \centering
    \includegraphics[width=0.9\linewidth]{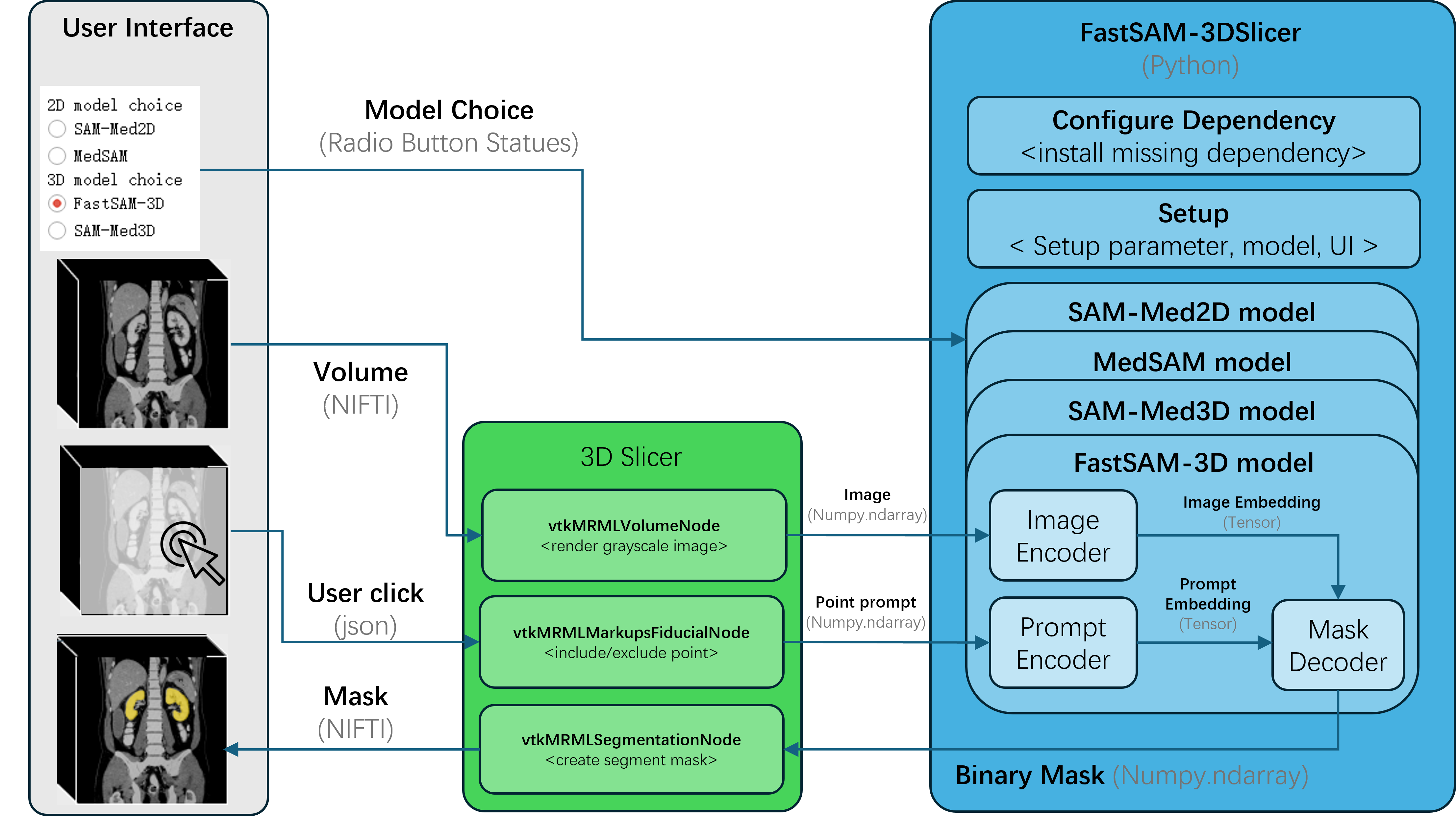}
    \caption{
The overall implementation workflow of FastSAM-3DSlicer. 
When FastSAM-3DSlicer is started for the first time, it runs the configure dependency block to check and install all required dependencies. 
The setup block runs every time FastSAM-3DSlicer is started to initialize parameters, models, and the user interface. 
Users can select a SAM model using a radio button. 
The \textcolor{WorkflowGreen}{green} blocks represent the components of the 3D Slicer environment, including the volume node, markup nodes, and segmentation node, which handle the raw image data, user interactions, and generated masks, respectively. 
The \textcolor{WorkflowBlue}{blue} blocks represent the FastSAM-3DSlicer Python components, which include configuring dependencies, setup, and the various supported 2D and 3D SAM models (SAM-Med2D \cite{sammed2d}, MedSAM \cite{medsam}, SAM-Med3D \cite{sammed3d}, and FastSAM-3D \cite{shen2024fastsam3d}). 
Black text indicates processes or actions within the workflow, while \textcolor{WorkflowGray}{gray} text represents data types or storage formats used throughout the process.
}\label{fig:structure}
\end{figure}

\section{Methods}
\subsubsection{Overview of the 3D Slicer Extension}
The interface for our FastSAM-3DSlicer extension is illustrated in Fig.~\ref{fig:interface}, with its overall implementation details shown in Fig.~\ref{fig:structure}.
Unlike previous works integrating 2D SAMs into 3D Slicer, such as TomoSAM \cite{tomosam} and SAMME \cite{samme}, FastSAM-3D's efficiency at the volume level eliminates the need to prepare image embeddings before interactive segmentation (\textit{i}.\textit{e}., user prompting) in 3D Slicer. 
This improvement enhances the user experience by reducing pre-processing time and allowing for more immediate interaction with the data.
Upon importing a 3D volumetric image into FastSAM-3DSlicer, the extension automatically generates three node types, including a volume node, a segmentation node, and two markup nodes.  
These node types represent specific data structures within 3D Slicer that manage different aspects of the image and segmentation process.
The volume node contains the raw 3D image in grayscale, represented as a NumPy array, which serves as the input for the SAM model
The segmentation node stores all the segmented masks, which are updated when users add new point prompts and generated when users perform the add mask operation.  
The two markup nodes hold all the include and exclude points that the user can add interactively.
When a user adds an include or exclude point prompt, FastSAM-3DSlicer first converts this point from RAS to XYZ coordinates based on the affine matrix stored in the input NIfTI file.  
Using this new point prompt and previous input point prompts, FastSAM-3DSlicer crops or pads the raw image to match the selected SAM model's input size and feeds it into the image encoder to generate image embeddings in real time.
The prompt encoder processes all the input points to generate prompt embeddings.
The mask decoder then translates the image embeddings, prompt embeddings, and previous masks into a new mask.
This segmentation mask is resized to the raw image dimensions based on saved coordinate information and updated in the segmentation node.
In addition to supporting the FastSAM-3D model, FastSAM-3DSlicer also supports SAM-Med3D \cite{sammed3d}, MedSAM \cite{medsam}, and SAM-Med2D \cite{sammed2d}, all of which follow the same structural process.


\subsubsection{Uncertainty Quantification Scheme}
Uncertainty quantification in FastSAM-3DSlicer provides users with a measure of confidence in the segmentation results, which can be used to guide user prompting. 
Regions with higher uncertainty indicate a greater need for additional prompts.
Our method leverages the efficiency of FastSAM-3D by running the image encoder once and performing multiple decoding steps to generate an ensemble of segmentations.
It begins with the initial segmentation, where the image encoder processes the input 3D volumetric image $\mathbf{I}$ to produce its image embedding $\mathbf{E} = \text{Encoder}(\mathbf{I})$. 
Next, the mask decoder generates the initial segmentation logits $\widehat{\mathbf{M}}_0$ based on the image embedding $\mathbf{E}$ and the initial set of point prompts ($\mathbf{P}_0$) provided from the user, \textit{i}.\textit{e}. $\widehat{\mathbf{M}}_0 = \text{Decoder}(\mathbf{E}, \mathbf{P}_0)$.
The segmentation mask $\mathbf{M}_0$ is the binarized segmentation logits $\widehat{\mathbf{M}}_0$, obtained by applying a threshold $\tau$:
\begin{equation}
\mathbf{M}_0 = \mathbbm{1}(\widehat{\mathbf{M}}_0 > \tau),
\end{equation}
where $\mathbbm{1}(\cdot)$ is the indicator function.
To quantify uncertainty, subsequent point prompts are sampled from the initial segmentation mask $\mathbf{M}_0$.  
These pseudo-point prompts are used to run the decoder multiple times, each time producing a slightly different segmentation mask due to variations in the sampled prompts. 
Let $\mathbf{P}_i$ denote the point prompts sampled from the initial segmentation mask $\mathbf{M}_0$ for the $i$\textsuperscript{th} iteration: $\mathbf{P}_i = \text{SamplePrompts}(\mathbf{M}_0)$.
The decoder is then run $N$ times using these sampled prompts, while keeping the image encoder constant, to produce $N$ different segmentation masks $\{\widehat{\mathbf{M}}_i\}_{i=1}^{N}$ with $\widehat{\mathbf{M}}_i = \text{Decoder}(\mathbf{E}, \mathbf{P}_i)$. 
Since the image encoder only runs once, the majority of the computational efficiency is preserved, as the decoder accounts for a smaller proportion of the total computation.
Inspired by the self-ensembling \cite{zhang2023segment}, the segmentation logits from each decoder run are averaged to produce the final ensemble result $\widehat{\mathbf{M}}$:
\begin{equation}
\widehat{\mathbf{M}} = \frac{1}{N} \sum_{i=1}^{N} \widehat{\mathbf{M}}_i.  
\end{equation}
This averaging process not only provides a robust final segmentation mask but also allows for the calculation of uncertainty. 
Specifically, the variability among the $N$ segmentation masks can be used to estimate the uncertainty, formally expressed as the standard deviation or variance of the logits at each voxel:
\begin{equation}
\text{Uncertainty}(x) = \sqrt{\frac{1}{N} \sum_{i=1}^{N} (\widehat{\mathbf{M}}_i(x) - \widehat{\mathbf{M}}(x))^2}
\end{equation}
where $x$ denotes a voxel in the 3D volume.

\section{Experiments}

\subsubsection{Implementation Details}
We implemented the proposed 3D Slicer extension using 3D Slicer version 5.6.2 and Python version 3.10.
The experiments were conducted in two distinct environments to evaluate inference times.
The first environment was a CPU-only setup utilizing a laptop-level AMD Ryzen 5 5500U CPU, while the second environment employed a GPU setup with one NVIDIA RTX 2060 GPU.
Following previous work \cite{shen2024fastsam3d}, we use the test set of \textit{totalsegmentator} \cite{totalsegmentator} to test the inference time.
All data are prepared in NIfTI format for loading.
Code for the 3D Slicer extension is available at \url{https://github.com/arcadelab/FastSAM3D_slicer}.
A video demo for our 3D slicer extension is available at \url{https://www.youtube.com/watch?v=oJ9ZhnPWqSs}. 

\subsubsection{Results for Inference Time}

\begin{table}[ht!]
\centering
\caption{Comparison of inference times for different SAMs on CPU and GPU in our 3D Slicer extension.
Results in \textcolor{myRed}{red} indicate slice-level inference times, while results in \textcolor{myBlue}{blue} indicate volume-level inference times.
}
\label{tab:inference_times}
\begin{tabular}{>{\raggedright\arraybackslash}m{2.5cm}| >{\centering\arraybackslash}m{4cm} >{\centering\arraybackslash}m{4cm}}
\toprule
\textbf{Models} & Inference Time w/ \textbf{CPU} & Inference Time w/ \textbf{GPU} \\
\midrule
SAM-Med2D \cite{sammed2d} & \textcolor{myRed}{1.52 seconds per slice} & \textcolor{myRed}{0.52 seconds per slice} \\
MedSAM \cite{medsam} & \textcolor{myRed}{48.9 seconds per slice} & \textcolor{myRed}{12.69 seconds per slice} \\
\hline
SAM-Med3D \cite{sammed3d} & \textcolor{myBlue}{7.75 seconds per volume} & \textcolor{myBlue}{1.76 seconds per volume} \\
FastSAM-3D \cite{shen2024fastsam3d} & \textcolor{myBlue}{1.09 seconds per volume} & \textcolor{myBlue}{0.73 seconds per volume} \\
\bottomrule
\end{tabular}
\end{table}

Table \ref{tab:inference_times} presents a comparison of inference times for different SAMs with both CPU and GPU environments within our FastSAM-3DSlicer extension. 
The SAM-Med2D \cite{sammed2d} exhibits an inference time of 1.52 seconds per slice on the CPU and 0.52 seconds per slice on the GPU.
While SAM-Med2D \cite{sammed2d} is effective for 2D slice-based segmentation, it fails to address volume-level segmentation in 3D Slicer, as the user needs to provide prompts for each slice individually, which results in longer inference time.
MedSAM \cite{medsam} shows considerably higher inference times, with 48.9 seconds per slice on the CPU and 12.69 seconds per slice on the GPU.
The increased processing time is due to its higher resolution of $1024\times1024$ compared to the $256\times256$ resolution of SAM-Med2D \cite{sammed2d}. 
Its high computational cost limits its practicality for real-time applications in image analysis settings.
The SAM-Med3D \cite{sammed3d}, optimized for 3D segmentation, achieves 7.75 seconds per volume on the CPU and 1.76 seconds per volume on the GPU.
This demonstrates a substantial improvement over MedSAM \cite{medsam}, particularly in GPU environments, making it a more viable option for volume-level segmentation.
FastSAM-3D \cite{shen2024fastsam3d} significantly outperforms the other models in terms of inference speed ($p<0.01$). 
It achieves 1.09 seconds per volume on the CPU and 0.73 seconds per volume on the GPU.
This reduction in inference time highlights the efficiency and optimization of FastSAM-3D for real-time interactive segmentation of 3D volumes.
FastSAM-3D demonstrates the lowest inference times on volume-level segmentation across both CPU and GPU environments, affirming its suitability for integration into image analysis workflows where speed and efficiency are critical. 
The improvements in inference time not only facilitate faster segmentation but also enable more immediate and iterative interaction with the volumetric data, thereby enhancing the overall utility of the 3D Slicer extension in medical image analysis applications.

\subsubsection{Illustrative Visual Examples}

\begin{figure}[t!]
    \centering
    \includegraphics[width=1.0\linewidth]{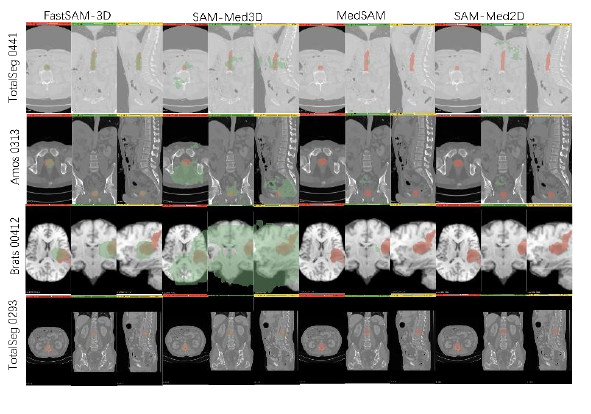}
    \caption{Qualitative comparison of segmentation results from different SAM models in the FastSAM-3DSlicer interface. Each row showcases the segmentation performance on a specific anatomical structure or region of interest. 
    The \textcolor{red}{red}  masks represent the ground truth segmentations, while the \textcolor{green}{green} masks depict the predicted segmentations from each model. 
    %
    }\label{fig:result}
\end{figure}


Fig. \ref{fig:result} presents a qualitative comparison of segmentation results obtained from different SAM models integrated into the FastSAM-3DSlicer interface. 
The examples showcase the performance of each model on various anatomical structures and regions of interest. 
%
%
FastSAM-3D and SAM-Med3D generate segmentations for the entire 3D volume, demonstrating their ability to capture spatial context and produce coherent masks. 
In contrast, MedSAM and SAM-Med2D operate on 2D slices, resulting in smaller and more localized segmentations when visualized in the 3D view. 
Across all examples, FastSAM-3D exhibits the highest agreement with the ground truth, highlighting its superior performance in terms of both efficiency and accuracy for real-time 3D interactive segmentation within the FastSAM-3DSlicer. 
%
%

\subsubsection{Results for Uncertainty Quantification}

\begin{figure}[t!]
    \centering
    \includegraphics[width=1.0\linewidth]{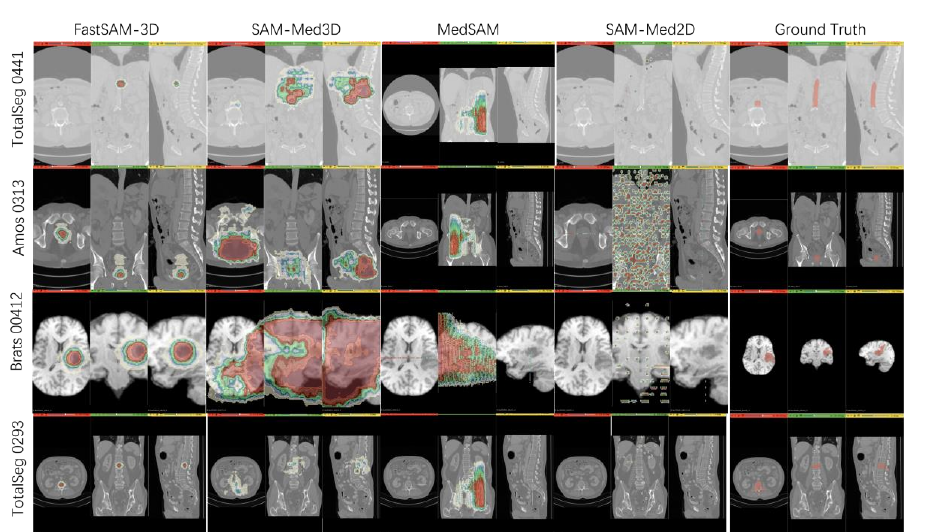}
    \caption{Visualization of the uncertainty quantification results obtained through self-ensembling in the FastSAM-3DSlicer.
    Each column represents a different SAM model, while the last column shows the ground truth segmentation. 
    The uncertainty maps are displayed as heatmaps overlaid on the original images, where lighter colors indicate higher uncertainty (lower probability) in the segmentation. 
    %
    %
    }\label{fig:selfensembling}
\end{figure}

Fig. \ref{fig:selfensembling} showcases the uncertainty quantification results obtained through the self-ensembling approach in the FastSAM-3DSlicer extension. 
%
%
%
The results demonstrate that FastSAM-3D exhibits the lowest overall uncertainty among the compared models, highlighting its robustness and reliability for real-time 3D interactive segmentation. 
%
%
It allows for the estimation of uncertainty by calculating the variance among the ensemble of segmentations.
The uncertainty information provided by FastSAM-3DSlicer is particularly valuable for guiding user interactions, as it identifies regions where additional user prompts may be required to improve segmentation accuracy. 
By focusing on areas of high uncertainty, users can iteratively refine the segmentation results, leading to more precise and reliable delineations of anatomical structures or regions of interest. 
%

\section{Conclusion}
We presented FastSAM-3DSlicer, a 3D Slicer extension designed to facilitate real-time, interactive segmentation of 3D volumetric medical images with SAMs. 
Our extension integrates SAM-Med2D, MedSAM, SAM-Med3D, and FastSAM-3D, offering a user-friendly interface that automates the handling of user prompts and segmented masks within the familiar 3D Slicer environment.
Our extension demonstrates superior performance in terms of inference time, particularly with the FastSAM-3D model, which achieves low inference times on both CPU and GPU environments. 
This makes FastSAM-3D highly suitable for real-time applications, reducing the computational burden while maintaining high segmentation quality. 
Furthermore, the integration of an innovative uncertainty quantification scheme leverages the rapid inference capabilities of FastSAM-3D, providing users with additional information about the reliability of the segmentation results.
Overall, by combining computational efficiency, precision, and ease of use, FastSAM-3DSlicer addresses the need for a user-friendly interface for SAMs, thereby enhancing decision-making processes and improving patient outcomes in medical settings.

\bibliographystyle{splncs04}
\bibliography{main.bib}

\begin{thebibliography}{10}
\providecommand{\url}[1]{\texttt{#1}}
\providecommand{\urlprefix}{URL }
\providecommand{\doi}[1]{https://doi.org/#1}

\bibitem{sammed2d}
Cheng, J., et~al.: Sam-med2d. arXiv preprint arXiv:2308.16184  (2023)

\bibitem{vit}
Dosovitskiy, A., et~al.: An image is worth 16x16 words: Transformers for image recognition at scale. arXiv preprint arXiv:2010.11929  (2020)

\bibitem{3dslicer}
Fedorov, A., et~al.: 3d slicer as an image computing platform for the quantitative imaging network. Magnetic resonance imaging  \textbf{30}(9),  1323--1341 (2012)

\bibitem{nnunet}
Isensee, F., et~al.: nnu-net: a self-configuring method for deep learning-based biomedical image segmentation. Nature methods  \textbf{18}(2),  203--211 (2021)

\bibitem{sam}
Kirillov, A., et~al.: Segment anything. arXiv preprint arXiv:2304.02643  (2023)

\bibitem{liu2021review}
Liu, X., et~al.: Review of deep learning based automatic segmentation for lung cancer radiotherapy. Frontiers in oncology  \textbf{11},  717039 (2021)

\bibitem{samme}
Liu, Y., et~al.: Segment any medical model extended. In: Medical Imaging 2024: Image Processing. vol. 12926, pp. 411--422. SPIE (2024)

\bibitem{medsam}
Ma, J., et~al.: Segment anything in medical images. Nature Communications  \textbf{15}(1), ~654 (2024)

\bibitem{tomosam}
Semeraro, F., et~al.: Tomosam: a 3d slicer extension using sam for tomography segmentation. arXiv preprint arXiv:2306.08609  (2023)

\bibitem{shen2024fastsam3d}
Shen, Y., et~al.: Fastsam3d: An efficient segment anything model for 3d volumetric medical images. arXiv preprint arXiv:2403.09827  (2024)

\bibitem{sammed3d}
Wang, H., et~al.: Sam-med3d. arXiv preprint arXiv:2310.15161  (2023)

\bibitem{totalsegmentator}
Wasserthal, J., et~al.: Totalsegmentator: Robust segmentation of 104 anatomic structures in ct images. Radiology: Artificial Intelligence  \textbf{5}(5) (2023)

\bibitem{zhang2023segment}
Zhang, Y., Hu, S., Jiang, C., Cheng, Y., Qi, Y.: Segment anything model with uncertainty rectification for auto-prompting medical image segmentation. arXiv preprint arXiv:2311.10529  (2023)

\end{thebibliography}

\end{document}